\documentstyle[twocolumn,aps,epsf]{revtex}

\tightenlines

\begin{document}

\draft

\title{Quantum limits on phase-shift detection using multimode interferometers}

\date{\today}

\author{Jonas S\"{o}derholm$^{1,2}$, Gunnar Bj\"{o}rk$^{1}$, Bj\"{o}rn Hessmo$^{1}$, and Shuichiro Inoue$^{3}$}

\address{$^{1}$Department of Microelectronics and Information Technology, Royal Institute of Technology (KTH), \\
Electrum 229, SE-164 40 Kista, Sweden \\
$^{2}$Department of Physics and Astronomy, University of New Mexico, Albuquerque, New Mexico 87131-1156, USA \\
$^{3}$Institute of Quantum Science, Nihon University, 1-8
Kanda-Surugadai, Chiyoda-ku, Tokyo 101-8308, Japan}

\maketitle

\begin{abstract}
Fundamental phase-shift detection properties of optical multimode 
interferometers are analyzed. Limits on perfectly distinguishable 
phase shifts are derived for general quantum states of a given 
average energy. In contrast to earlier work, the limits are found 
to be independent of the number of interfering modes. However, 
the reported bounds are consistent with the Heisenberg limit. A 
short discussion on the concept of well-defined relative phase is 
also included.
\end{abstract}

\pacs{PACS numbers: 42.50.-p, 03.65.Ta, 07.60.Ly, 42.50.Dv}

\section{Introduction}

It is today well-known that the use of quantum-mechanical states 
can improve the precision of interferometric measurements. 
According to the so-called standard quantum limit \cite{CMC}, the 
precision of optical measurements employing classical states 
cannot increase faster than $\propto 1/\sqrt{E}$, where $E$ is 
the energy used in the measurement. However, the use of 
nonclassical states allows us to reach the Heisenberg limit 
$\propto 1/E$ \cite{Ou}, which for high energies would give a 
remarkable improvement in accuracy. Several setups have 
theoretically been shown to work at the Heisenberg limit 
\cite{Caves,Holland,PRL,Bollinger}. The standard quantum limit 
has also been circumvented experimentally 
\cite{Kimble,Grangier,Heidmann,Polzik}. However, the fragile 
nature of the quantum states has so far prevented these 
measurements from being carried out with higher energies. 
Therefore, high intensity classical interferometry reach a much 
better overall accuracy.

Recently, a bound closely related to the Heisenberg limit was 
given by Margolus and Levitin \cite{Margolus}. The bound gives 
the time necessary for a state of a closed system to become 
orthogonal for a given average energy. This limits the rate of 
operations in quantum information processing and the resolution 
in interferometry. In an earlier paper \cite{PRA}, we derived the 
states that minimize the time needed to freely evolve into 
another state, whose overlap with the original one was given. 
This would correspond to minimizing the necessary phase shift for 
a single-mode state.

The present paper is devoted to the phase-shift detection 
properties of multimode interferometers. In many interferometric 
measurements, a single induced phase shift is tracked by 
monitoring the interference fringes in the output of a two-mode 
interferometer. However, it is often possible to induce several 
phase shifts of the same, or smaller, magnitude (possibly with 
different signs) in the different arms of a multimode 
interferometer. Here, we investigate whether this fact can be 
used to improve the accuracy of such measurements. We will assume 
that the interfering fields have the same optical frequency, so 
that the total energy is proportional to the number of photons 
used.

In an earlier investigation \cite{DAriano}, it was concluded that 
the accuracy of the multimode interferometer would improve 
indefinitely with the number of modes. However, we find that 
there is no fundamental advantage in using more than two modes 
and that, in our eyes, the conclusions in Ref.\ \cite{DAriano} 
stem from an unfortunate choice of figure of merit. Rather, the 
accuracy is found to be limited only by the energy used in the 
measurement and scale according to the Heisenberg limit.

\section{Problem formulation}

Accuracy can be defined in many different ways. For example, in a 
recent paper on multimode interferometry \cite{Sanders}, the 
width of the major peak of the phase distribution was taken as a 
measure of the precision. Here, we will define it as the smallest 
phase shift required to give a perfectly distinguishable outcome, 
i.e., the phase-shifted state is required to be orthogonal to the 
original state.

\begin{figure}[htbp]
  \begin{center}\hspace{0mm}\mbox{\input epsf
\epsfxsize7.0cm\epsfbox{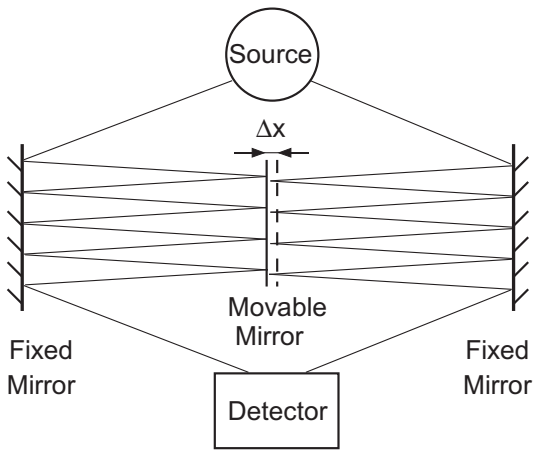}}\end{center}
  \caption{A sketch of an interferometric setup to probe the
  displacement $\Delta x$ of the two-sided mirror at the center.
  As the movable mirror is displaced by $\Delta x$ to the right,
  the phase shift in the left mode becomes $\phi_{\rm left}
  = 2 \pi R_{\rm left} \Delta x/\lambda_0 \cos \gamma$, where
  $R_{\rm left}$ is the number of times the beam runs between
  the left and center mirror, $\lambda_0$ is the optical wavelength,
  $\gamma$ is the angle of incidence, and we have assumed
  unit refractive index. The phase shift in the right mode
  is given by a similar expression, but has opposite sign. In
  practice, the induced phase shift is always limited, e.g., due
  to the quality of the light beam. However, note that we, in
  principle, can use as many modes on each side of the movable
  mirror as we like. By an appropriate choice of the number of
  reflections and the angle of incidence for mode $m$, we can
  realize any phase shift satisfying $|\phi_m| \leq \phi$,
  where $\phi$ is the greatest possible phase shift for a single
  mode.}
  \label{fig:setup}
\end{figure}

\begin{figure}[htbp]
  \begin{center}\hspace{0mm}\mbox{\input epsf
\epsfxsize7.0cm\epsfbox{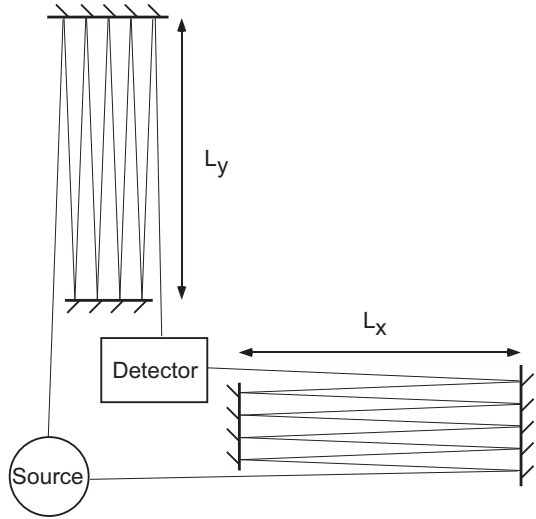}}\end{center}
  \caption{A generalized delay-line Michelson interferometer for
  detection of gravitational waves. As the $x$- and $y$-arm are
  perpendicular, quadrupole gravitational waves would, under
  certain conditions, induce changes in the mirror separations
  $L_x$ and $L_y$ that are of the same magnitude but with
  different signs [15]. The present setup then becomes
  equivalent to the one sketched in Fig.\ \ref{fig:setup}.
  Note that we consider a general source and detector, whereas
  the conventional delay-line Michelson interferometer used for
  gravitational wave detection consists of a laser source, a
  beam splitter and a photodetector.}
  \label{fig:grav}
\end{figure}

More precisely, we look for the smallest phase shift that can be 
detected with certainty using an $M$-mode interferometer and a 
state whose average energy is given. The smallest phase shift 
$\phi$ should here be interpreted in the following sense (cf. 
Figs.\ \ref{fig:setup} and \ref{fig:grav}). We consider the case 
where all arms of the interferometer have induced phase shifts 
that satisfy $\phi_m = \lambda_m \phi$, where $-1 \leq \lambda_m 
\leq 1$ and $\phi > 0$ is the parameter to be minimized under the 
condition that it results in a state that is perfectly 
distinguishable from the original. Since the $N$-photon state 
$(|0,N \rangle + |N,0 \rangle )/\sqrt{2}$ can be made orthogonal 
in a two-mode interferometer by a relative phase shift of $\pi/N$ 
\cite{PRL}, and the two-mode interferometer is a special case of 
the multimode interferometer considered here, we know that the 
smallest necessary phase shift $\phi$ is smaller than, or equal 
to, $\pi/2 N$.

Minimizing $\phi$ above corresponds to an experimental situation 
where we have an arbitrary number of induced phase shifts at our 
disposal (one in each mode). However, we easily identify another 
experimental situation of interest, where we have a given amount 
of phase shift that can be split and distributed among the modes. 
If we are also allowed to choose the signs of the phase shifts, 
our problem is to minimize $\phi_{\rm tot} = \sum_{m=1}^M 
|\phi_m|$. We will see that the solution to this problem also 
gives a tight bound for the case where all phase shifts have the 
same sign.

In accordance with our discussion above, we consider an $M$-mode 
interferometer with phase shifts $\phi_m = \lambda_m \phi$, where 
$-1 \leq \lambda_m \leq 1$, in the different arms. We will also, 
without any loss of generality, number the modes in such a way 
that $\lambda_1 \geq \lambda_2 \geq \ldots \geq \lambda_M$. The 
two problems formulated above can then be expressed in a 
mathematical form as follows: We want to find the smallest value 
of $\phi$ and $\phi_{\rm tot} = \sum_{m=1}^M |\phi_m|$, 
respectively, such that \begin{equation} \langle \Psi |e^{i \phi 
\sum_{m=1}^M \lambda_m \hat{n}_m}| \Psi \rangle = 0 , 
\label{eq:condition} 
\end{equation} for some pure $M$-mode state $| \Psi \rangle$ with 
a given average energy. In particular, we are, in correspondence 
with the Heisenberg limit, looking for the minimum product of our 
accuracy measure ($\phi$ or $\phi_{\rm tot}$) and the average 
photon number $\langle \hat{N} \rangle$.

\section{The energy eigenstates} \label{sec:energy}

In this section, we restrict ourself to the $N$-photon states. We 
derive lower bounds on $\phi$ and $\phi_{\rm tot}$, and give 
examples that saturate these bounds. The general case, where the 
states are allowed to be superpositions of different photon 
number states, is considered in Sec.\ \ref{sec:general}.

\subsection{Minimizing $\phi$ with respect to N}

Let $| \Psi_N \rangle$ denote a pure $M$-mode, $N$-photon state 
and introduce the nonnegative parameters $\mu_m = \lambda_m - 
\lambda_M$. The orthogonality condition (\ref{eq:condition}) then 
becomes \begin{equation} \langle \Psi_N |e^{i \phi \sum_{m=1}^M 
\mu_m \hat{n}_m}| \Psi_N \rangle = 0 , \label{eq:condition2} 
\end{equation} where $0 \leq \mu_m \leq 2$ and $\mu_1 \geq \mu_2 
\geq \ldots \geq \mu_M = 0$. Furthermore, let $\{ | {\cal 
N}_j^{(N)} \rangle \}_{j=1}^{J_N}$ denote the set of all possible 
orthogonal states \begin{equation} | {\cal N}_j^{(N)} \rangle = 
|n_{1,j}^{(N)},n_{2,j}^{(N)},\ldots, n_{M,j}^{(N)} \rangle 
\end{equation} that for each $j$ fulfill \begin{equation} \sum_{m=1}^{M} 
n_{m,j}^{(N)} = N , \quad \quad  n_{m,j}^{(N)} \in \{ 
0,1,2,\ldots,N \} . \end{equation} (The superscripts are not 
necessary here, but will be used in the next section.) A general 
pure $N$-photon state can then be written \begin{equation} | 
\Psi_N \rangle = \sum_{j=1}^{J_N} c_j^{(N)} | {\cal N}_j^{(N)} 
\rangle , \label{eq:PsiN} \end{equation} where $\sum_{j=1}^{J_N}  
|c_j^{(N)}|^2 = 1$. Putting this into Eq.\ (\ref{eq:condition2}) 
gives \begin{eqnarray} \lefteqn{\sum_{j=1}^{J_N} \left| c_j^{(N)} 
\right|^2 \cos \left( \phi \sum_{m=1}^{M} \mu_m n_{m,j}^{(N)}
\right)} & & \nonumber \\
& & + i \sum_{j=1}^{J_N} \left| c_j^{(N)} \right|^2 \sin \left( 
\phi \sum_{m=1}^{M} \mu_m n_{m,j}^{(N)} \right) = 0 . 
\label{eq:cond3} 
\end{eqnarray} In order for the imaginary part to vanish, the 
smallest definite phase shift we can use is $\phi = \pi/2 N$, 
which implies that $c_j^{(N)} = 0$, for all $j$ such that 
$\sum_{m=1}^M \mu_m n_{m,j}^{(N)} \neq 0$ or $2 N$. We must also 
have $\mu_1 = 2$ (remember that the parameters are labeled in 
decreasing order). Now, let $p_0$ denote the sum of 
$|c_j^{(N)}|^2$ over all $j$ such that $\sum_{m=1}^M \mu_m 
n_{m,j}^{(N)} = 0$. The condition for the real part to vanish 
then becomes $p_0 - (1 - p_0) = 0$, i.e., $p_0 = 1/2$. The 
simplest state that fulfills this has only two nonzero 
coefficients $c_j^{(N)}$. An example of such a state is 
\begin{equation} | \Phi_N \rangle = \frac{1}{\sqrt{2}} 
\left( |N,0,\ldots,0 \rangle + e^{i \vartheta} |0,0,\ldots,N 
\rangle \right) , \label{eq:PhiN} \end{equation} where 
$\vartheta$ is an arbitrary real number. This state makes it 
possible to detect a phase shift of $\pi/2 N$ with certainty. As 
shown above, this is the smallest phase shift that can be 
detected with certainty using $N$-photon states. We also note 
that the state (\ref{eq:PhiN}) can be seen as a representation of 
a two-mode interferometer, since all but two modes (arms) are 
left unexcited. That is, the resolution limit $\phi = \pi/2 N$ 
can be reached with a two-mode interferometer.

\subsection{Minimizing $\phi_{\rm tot}$ with respect to N}

Now, assume that $| \Psi_N \rangle$ is an $N$-photon state that 
together with the phase shifts $\phi_m = \lambda_m \phi$ minimize 
$\phi_{\rm tot} = \sum_{m=1}^M |\phi_m|$ under the condition 
(\ref{eq:condition}). Equation (\ref{eq:condition}) then implies 
that Eqs.\ (\ref{eq:condition2}) and (\ref{eq:cond3}) are 
satisfied, where $\mu_m \phi = \phi_m - \phi_M$. Since the real 
part of the expression (\ref{eq:cond3}) must vanish, we have 
$\phi \sum_{m=1}^{M} \mu_m n_{m,j}^{(N)} > 0$, for some $j$. In 
order for the imaginary part to vanish, it then follows that 
$\phi \sum_{m=1}^{M} \mu_m n_{m,j}^{(N)} \geq \pi$, for some $j$. 
Thus, there exists an $m$ such that $\phi_m - \phi_M \geq \pi/N$, 
i.e., $\phi_{\rm tot} \geq \pi/N$. This bound, too, can be 
saturated by the state (\ref{eq:PhiN}), if we choose $\phi_1 = 
\pi/2 N$ and $\phi_M = - \pi/2 N$. Again, we find that the lower 
bound can be achieved with a two-mode interferometer.

\subsection{The de Broglie wavelength}

Note that the bounds we have obtained for the $N$-photon states 
are easily interpreted in terms of their de Broglie wavelength 
$\lambda_{\rm dB} = \lambda_0/N$, where $\lambda_0$ is the 
optical wavelength \cite{PRL}. We know that an ordinary two-mode 
interferometer fed with classical light (or a single photon) 
needs a relative phase shift of $\pi$, corresponding to half the 
optical wavelength, in order to change between constructive and 
destructive interference. If we instead feed the interferometer 
with nonclassical states of light with an average energy $\langle 
\hat{N} \rangle$, the corresponding de Broglie wavelength is 
limited from below by $\lambda_0/\langle \hat{N} \rangle$. Thus, 
the necessary total phase shift is $\phi_{\rm tot} = \pi/\langle 
\hat{N} \rangle$. That the bound does not depend of the number of 
modes $M$, can be seen as a consequence of the fact that the de 
Broglie wavelength corresponds to the interfering entities 
(``photon clusters") and not to the number of interfering paths.  
However, one could expect that it would be possible to divide 
this total phase shift equally among all the $M$ modes, so that 
$\phi = \pi/M \langle \hat{N} \rangle$ would be sufficient. 
However, as we have seen above, this is not true. We found that 
the total phase shift can only be divided between two modes, 
which gives $\phi \geq \pi/2 \langle \hat{N} \rangle$ and makes 
it possible to reach the bound with a two-mode interferometer.

To simplify comparisons between the limits for different 
experimental situations, we have gathered our results in Table 
\ref{table:ll}. The bounds for the single-mode case follow from 
Ref.\ \cite{Margolus}, and the set of states saturating these was 
derived in Ref.\ \cite{PRA}. Since single-mode photon number 
states cannot be made orthogonal by a phase shift, the bounds 
found in this section are the only limits given for the 
$N$-photon states in the table. 

\section{The general case} \label{sec:general}

Next, we tackle our problem for a general pure state. We make 
some general considerations before we concentrate on the 
minimization of the respective measures $\phi$ and $\phi_{\rm 
tot}$ in two subsections.

\begin{table}
\caption{Lower limits on the distinguishable phase shifts.}
\begin{tabular}{l c c c} \label{table:ll}
& Entity & One mode & Two or more modes \\
\hline Arbitrary states & $\phi$ & $\pi/2 \langle \hat{N} 
\rangle$ & $\approx 1.38005/\langle \hat{N} \rangle$ \\
& $\phi_{\rm tot}$ & $\pi/2 \langle \hat{N} \rangle$ & $\pi/2
\langle \hat{N} \rangle$ \\
\hline
$N$-photon states & $\phi$ & NA & $\pi/2 \langle \hat{N} \rangle$ \\
& $\phi_{\rm tot}$ & NA & $\pi/\langle \hat{N} \rangle$ \\
\end{tabular}
\end{table}

Let us start by denoting a general pure state as \begin{equation} 
| \Psi \rangle = \sum_{N=0}^\infty r_N e^{i \varphi_N} | \Psi_N 
\rangle , \label{eq:superpos} \end{equation} where $r_N$ and 
$\varphi_N$ are real and nonnegative, $\sum_{N=0}^\infty r_N^2 = 
1$, and the states $| \Psi_N \rangle$ are given by Eq.\ 
(\ref{eq:PsiN}). Our orthogonality condition (\ref{eq:condition}) 
can then be written 
\begin{equation} \sum_{N=0}^\infty r_N^2 \sigma_N = 0 , \label{eq:ortho2} 
\end{equation} where we have defined \begin{equation} \sigma_N \equiv \sum_{j=1}^{J_N} 
\left| c_j^{(N)} \right|^2 \exp \left( i \phi \sum_{m=1}^{M} 
\lambda_m n_{m,j}^{(N)} \right) . \label{eq:sigmaN} \end{equation}

The average photon number of the state (\ref{eq:superpos}) is 
\begin{equation} \langle \hat{N} \rangle = \sum_{N=0}^\infty 
r_N^2 N . \label{eq:meanN} \end{equation} In order to find the 
normalized states that minimize the energy for a given $\phi$ 
under the condition (\ref{eq:ortho2}), we define the function 
\begin{equation} F \equiv \sum_{N=0}^\infty r_N^2 N + \alpha 
\left( 1 - \sum_{N=0}^\infty r_N^2 \right) + \beta 
\sum_{N=0}^\infty r_N^2 \sigma_N , \end{equation} where $\alpha$ 
and $\beta$ are Lagrange's multipliers. Any state that minimizes 
the energy represents an extremum of $F$. Therefore, these states 
must satisfy 
\begin{eqnarray} \frac{\partial F}{\partial r_k} & = & 2 r_k 
\left( k - \alpha + \beta \sigma_k \right) = 0 , \quad \forall k 
. \label{eq:optcond} 
\end{eqnarray} Since $k$, $\alpha$, and $\beta$ are real, this 
means that for each $k$ such that $r_k \neq 0$, $\sigma_k$ must 
be real.

Now, let the values of $k$ satisfying $r_k \neq 0$ be denoted 
$k_q$, and let $| \Theta \rangle = \sum_{q} r_{k_q} \exp (i 
\varphi_{k_q}) | \Psi_{k_q} \rangle$ be an optimal state. Then 
there always exists a largest positive value of $(1 - 
\sigma_{k_q})/k_q$ for those $k_q$ that satisfy $k_q \neq 0$. Let 
$N > 0$ denote one excited manifold that attains that value, and 
define the state 
\begin{equation} | \tilde{\Theta} \rangle = \sqrt{1 - r_N^2 - \delta} \, 
|0,0,\ldots,0 \rangle + \sqrt{r_N^2 + \delta} \, | \Psi_N \rangle 
, \label{eq:twoN} \end{equation} where 
\begin{equation} \delta = \sum_{\{ q | k_q \neq N \}} r_{k_q}^2 
\frac{1 - \sigma_{k_q}}{1 - \sigma_N} . 
\end{equation} It is easily verified that $| \tilde{\Theta} \rangle$ is 
normalized and satisfies the orthogonality condition 
(\ref{eq:ortho2}). The average photon number is found to be 
\begin{equation} r_N^2 N + \sum_{\{ q | k_q \neq N \}} r_{k_q}^2 
\frac{1 - \sigma_{k_q}}{1 - \sigma_N} N \leq \sum_q r_{k_q}^2 k_q 
, \label{eq:valid} \end{equation} i.e., it is smaller than, or 
equal to, the average photon number of $| \Theta \rangle$. Hence, 
all optimal combinations of the average photon number and $\phi$, 
or $\phi_{\rm tot}$, can be attained with states that are 
superpositions of the vacuum and one excited manifold $N$.

\subsection{Minimizing $\phi$ with respect to $\langle \hat{N} \rangle$}
\label{sec:genphi}

In accordance with our findings above, we restrict our attention, 
without loss of generality, to states of the form 
(\ref{eq:superpos}), which we rewrite as \begin{equation} | 
\tilde{\Theta} \rangle = r_0^2 | 0,0,\ldots,0 \rangle + \sqrt{1 - 
r_0^2} | \Psi_N \rangle . \label{eq:twoN2} \end{equation} From 
the orthogonality condition (\ref{eq:ortho2}) and Eq.\ 
(\ref{eq:meanN}), we then obtain $\sigma_N = - r_0^2/(1 - r_0^2) 
= 1 - N/\langle \hat{N} \rangle$. Thus, our problem is to 
minimize $\sigma_N$, and thereby $\langle \hat{N} \rangle$, for a 
given phase shift $\phi$. Assuming an optimal state, $\sigma_N$ 
must be real according to Eq.\ (\ref{eq:optcond}). It then 
follows from Eq.\ (\ref{eq:sigmaN}) that $\sigma_N \geq \cos (N 
\phi)$ for phase shifts $\phi \leq \pi/N$. The lower bound can be 
saturated, e.g., by choosing $\lambda_1 = 1$, $\lambda_M = -1$, 
and the state (\ref{eq:PhiN}). The minimum value $\sigma_N = -1$ 
is then reached with $\phi = \pi/N$. Introducing $\eta = N 
/\langle \hat{N} \rangle$, we have 
\begin{equation} \cos (N \phi) = 1 - \eta \quad \Rightarrow \quad 
\phi = \frac{\arccos (1 - \eta)}{\eta \langle \hat{N} \rangle} . 
\end{equation} In order for the phase-shifted and original state 
to be orthogonal, $r_0^2 \leq 1/2$ must be satisfied. Thus, we 
have $1 \leq \eta \leq 2$. The value of $\eta$ that minimizes 
$\phi$ for a given energy must satisfy \begin{equation} 
\frac{\partial \phi}{\partial \eta} = \frac{1}{\langle \hat{N} 
\rangle} \left[ \frac{1}{\eta \sqrt{1 - (1 - \eta)^2}} - 
\frac{\arccos (1 - \eta)}{\eta^2} \right] = 0 , 
\end{equation} which implies that
\begin{equation} \arccos (1 - \eta) = \sqrt{\frac{\eta}{2 - 
\eta}} . \end{equation} Thus, the optimal value of $\eta$ is 
found to be \begin{equation} \eta_{\rm opt} \approx 1.6891577 , 
\end{equation} which gives \begin{equation} \phi = \frac{\arccos 
(1 - \eta_{\rm opt})}{\eta_{\rm opt} \langle \hat{N} \rangle} 
\approx \frac{1.38005}{\langle \hat{N} \rangle} . 
\label{eq:bound} \end{equation} We notice that the bound can only 
be saturated for $\langle \hat{N} \rangle = N/\eta_{\rm opt}$, 
where $N$ is a positive integer. This can be achieved by 
choosing, e.g., $\lambda_1 = 1$, $\lambda_M = -1$, and the state 
\begin{eqnarray} | \Upsilon_N  \rangle & = & \sqrt{\frac{\eta_{\rm opt} - 
1}{\eta_{\rm opt}}} \, |0,0,\ldots,0 \rangle \nonumber \\
& & + \frac{1}{\sqrt{2 \eta_{\rm opt}}} \left( |N,0,\ldots,0 
\rangle + e^{i \vartheta} |0,0,\ldots,N \rangle \right) \nonumber \\
& \approx & 0.638740 \, |0,0,\ldots,0 \rangle + 0.769423 \, | 
\Phi_N \rangle , \label{eq:optst} \end{eqnarray} where 
$\vartheta$ is an arbitrary real number and $| \Phi_N \rangle$ is 
defined in Eq.\ (\ref{eq:PhiN}). Again, we see that two modes are 
sufficient to obtain the best possible accuracy.

\subsection{Minimizing $\phi_{\rm tot}$  with respect to $\langle \hat{N} \rangle$}
\label{sec:genphitot}

The minimum value of $\phi_{\rm tot}$ for a general state of a 
given average energy follows directly from the work by Margolus 
and Levitin \cite{Margolus}. They derived the time of free 
evolution necessary for a state to evolve into an orthogonal 
state. For the optical field, the necessary time is $T = 1/4 f 
\langle \hat{N} \rangle$, where $f$ is the optical frequency. 
Since this corresponds to a phase shift $\phi = \pi/2 \langle 
\hat{N} \rangle$, we must have $\phi_{\rm tot} \geq \pi/2 \langle 
\hat{N} \rangle$. The bound can be saturated, e.g., by using the 
state \begin{equation} | \Omega_N \rangle = \frac{1}{\sqrt{2}} 
\left( |0,0,\ldots,0 \rangle + |N,0,\ldots,0 \rangle \right) 
\label{eq:OmegaN} \end{equation} and choosing $\phi_1 = \pi/N$ as 
the only phase shift. We see that this bound, too, can be 
achieved with a two-mode interferometer. In fact, it can even be 
achieved with a single mode. For further comparisons, we refer to 
Table~\ref{table:ll}.

\section{Energy measurements and well-defined relative phase} 
\label{sec:wdrp}

\begin{figure}[htbp]
  \begin{center}\hspace{0mm}\mbox{\input epsf
\epsfxsize7.0cm\epsfbox{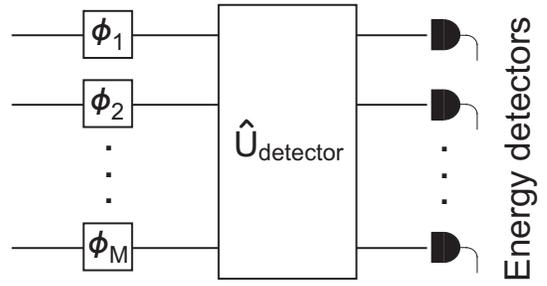}}\end{center}
  \caption{A schematic diagram of an interferometric measurement
  that also gives the total energy of the measured state. The
  interference of the $M$ modes is characterized by the lossless
  and unitary transformation $\hat{U}_{\rm detector}$.}
  \label{fig:measurement}
\end{figure}

In the previous sections, we have derived the smallest 
distinguishable phase shifts under two different restraints. We 
know from quantum theory that each time we make a measurement 
corresponding to a Hermitian operator with the original and 
phase-shifted state as eigenstates, the outcome will tell us 
whether the phase shift was applied or not. However, it is, in 
general, hard to realize such a measurement. In fact, most often 
the energy is measured. For example, in quantum optics it is 
usually the photon number, or the intensity of the optical field, 
that is finally probed. In Fig.\ \ref{fig:measurement}, we have 
drawn a general measurement setup of this kind. Such measurements 
would, in principle, allow us to achieve the resolution derived 
for the energy eigenstates in Sec.\ \ref{sec:energy}. In 
contrast, the vacuum component of the state (\ref{eq:optst}) is 
invariant under phase shifts, and makes it impossible to tell if 
the state has been phase-shifted or not when no energy is found 
in the measurement.

Ideally, the interference, which is represented by $\hat{U}_{\rm 
detector}$ in Fig.\ \ref{fig:measurement}, is lossless and 
unitary. In a measurement that gives the total energy, the 
requirement to be able to distinguish a phase shift $\phi$ with 
certainty can therefore be written \begin{equation} \langle \Psi 
| \hat{\openone}_N e^{i \phi \sum_{m=1}^M \lambda_m \hat{n}_m} | 
\Psi \rangle = 0 , \label{eq:distingM} \end{equation} where we 
have used the definition \begin{equation} \hat{\openone}_N \equiv 
\sum_{j=1}^{J_N} |{\cal N}_j^{(N)} \rangle \langle {\cal 
N}_j^{(N)}| . \label{eq:identityN} \end{equation} That is, for 
any possible outcome of the total energy, the original and 
phase-shifted states must be orthogonal.

In an earlier paper \cite{Inoue}, we claimed that if $| \xi 
\rangle$ is a two-mode state, $\hat{U}_{\rm PS} (\phi) = \exp(i 
\phi [\hat{n}_1 - \hat{n}_2]/2)$ is  the two-mode unitary 
relative-phase shifting operator, and if
\begin{equation}
\langle \xi | \hat{U}_{\rm PS} (\phi) | \xi \rangle = 0 
\label{eq:disting}
\end{equation}
can be fulfilled for some value of $\phi$, then ``Eq.\ 
(\ref{eq:disting}) constitutes the mathematical criterion for a 
state with a well-defined relative phase." It is clear that Eq.\ 
(\ref{eq:disting}) is a special case of Eq.\ 
(\ref{eq:condition}), and therefore constitutes the necessary and 
sufficient condition for a relative phase shift to be 
distinguishable. However, as the following example shows, this 
property should not be referred to as ``well-defined relative 
phase." Suppose that \begin{equation} | \xi \rangle = 
\frac{1}{\sqrt{2}} (|0,0 \rangle + |0,1 \rangle) = |0 \rangle 
\otimes \frac{1}{\sqrt{2}} (|0 \rangle + |1 \rangle) . 
\label{eq:xi} 
\end{equation} It is trivial to show that in this case Eq.\ 
(\ref{eq:disting}) is fulfilled for $\phi = 2 \pi$. However, a 
product state where one of the constituent factors is a single 
mode vacuum state is hardly a sensible candidate for a state with 
a well-defined relative phase. This is confirmed by the 
relative-phase distribution functions, which are flat \cite{PS}. 
The conclusion is that the criterion for an operationally 
well-defined relative phase must be reformulated. 

The constant relative-phase distribution functions of the state 
(\ref{eq:xi}) are a consequence of the fact that the 
relative-phase and the total energy are compatible observables 
\cite{LSS}. Since $\hat{U}_{\rm PS} (\phi)$ also commutes with 
the total photon number operator $\sum_{N=0}^\infty N 
\hat{\openone}_N$, any measurement of two-mode relative phase (or 
visibility) will reveal the total photon number. Therefore, in 
order for the phase shift $\phi$ to be perfectly distinguishable, 
the phase-shifted state $\hat{U}_{\rm PS}(\phi) | \xi \rangle$ 
must, as in Eq.\ (\ref{eq:distingM}), become orthogonal to the 
original state $| \xi \rangle$ in every photon-number manifold. 
Consequently, the definition of well-defined relative phase 
should be refined as follows. If, for some phase shift $\phi$ and 
for all $N$, it is possible to fulfill
\begin{equation}
\langle \xi | \hat{\openone}_N \hat{U}_{\rm PS} (\phi) | \xi 
\rangle = 0, \label{eq:disting 2}
\end{equation}
then the state $| \xi \rangle$ is said to have a well-defined 
relative phase. That any state fulfilling Eq.\ (\ref{eq:disting 
2}) also fulfills Eq.\ (\ref{eq:disting}) can easily be verified. 
However, the converse is not true, as shown by the state 
(\ref{eq:xi}), which does not fulfill Eq.\ (\ref{eq:disting 2}).

Since all the components in different excitation manifolds must 
be made orthogonal simultaneously, the limits found in Sec.\ 
\ref{sec:energy} imply that in this case $\phi \geq \pi/2 N_{\rm 
max}$ and $\phi_{\rm tot} \geq \pi/N_{\rm max}$, where $N_{\rm 
max}$ denotes the highest excited manifold.

\section{Discussion}

We have found tight bounds set by the average energy of the used
states on our accuracy measures $\phi$ and $\phi_{\rm tot}$. The 
bounds were derived for optical interferometers and pure states, 
but are obviously valid for other interfering bosons and mixed 
states, too. Our results also give tight bounds in the case where 
all phase shifts have the same sign. This would correspond to an 
experimental situation where there is a limited amount of 
phase-shift inducing material that can be distributed among the 
modes. Since the state (\ref{eq:PhiN}) can saturate the bounds 
for $\phi_{\rm tot}$ with nonnegative or nonpositive phase shifts 
$\phi_m$ alone, the magnitude of the sum of these phase shifts 
has the same bounds for $N$-photon states and general states as 
$\phi_{\rm tot}$.

We would also like to point out that our results were derived 
from fundamental considerations. Apart from the brief discussion 
in Sec.\ \ref{sec:wdrp}, we did not consider the problem of how 
to realize the measurements necessary to distinguish the optimal 
states. Although we have found that there is no fundamental 
advantage in using multimode interferometers in terms of their 
accuracy, they may offer simpler realizations compared to 
two-mode interferometers. For example, it may be easier to 
generate multimode states that give the same resolution as the 
two-mode states presented here.

Let us also comment on the earlier investigation of multimode 
interferometers by D'Ariano and Paris \cite{DAriano}, which 
arrived at the conclusion that the multimode interferometers 
actually improve the accuracy. They considered interferometers 
whose phase shifts in the different modes were $\phi_m = m 
\varphi$. The ``variance" of the estimate of $\varphi$ was found 
to be inversely proportional to the number of modes $M$, 
indicating that the accuracy in the measurement of $\varphi$ was 
improved correspondingly. However, we note that the largest 
relative phase shift $\phi_M - \phi_1 = (M - 1) \varphi$ 
increases with the number of modes $M$. Since the two-mode 
interferometer employs a relative phase shift of $\varphi$ 
instead of $(M - 1) \varphi$, its sensitivity is not maximized. 
The relative phase shift $(M - 1) \varphi$ would obviously give 
the same improvement in the accuracy of $\varphi$ for a two-mode 
interferometer as D'Ariano and Paris found for the $M$-mode 
interferometer. Therefore, we conclude that for a given average 
energy the ultimate phase-shift resolution of an interferometer 
is independent of the number of interfering modes. 

\acknowledgments

This work was funded by the Swedish Research Council. JS 
acknowledges support from John och Karin Engbloms Stipendiefond.

\end{document}